\documentstyle[12pt]{article}

\begin{document}

\begin{titlepage}

\hspace{9.5cm}{IFT-P.021/99}

\vspace{.5cm} 

\begin{center}

\LARGE

{\sc A Note on the Superstring BRST Operator}

\vspace{.5cm}
\large

Jos\'e N. Acosta\footnote{e-mail:  jose@ift.unesp.br},\\
Nathan Berkovits\footnote{e-mail:  nberkovi@ift.unesp.br}, \\ and \\ 
Osvaldo Chand\'{\i}a\footnote{e-mail:  chandia@ift.unesp.br}

\vspace{.5cm}

{\em Instituto de F\'\i sica Te\'orica, Universidade Estadual Paulista} \\ 
{\em Rua Pamplona 145, 01405-900, S\~ao Paulo, SP, Brasil}

\vspace{.5cm}

February, 1999

\end{center}

\vspace{1cm}

\begin{abstract}

We write the BRST operator of the N=1 superstring as
$Q= e^{-R} (\frac{1}{2\pi i}\oint dz ~
\gamma^2 b )e^R$ where $\gamma$ and $b$ are 
super-reparameterization ghosts. This provides a trivial
proof that $Q$ is nilpotent. 

\end{abstract}

\end{titlepage}

\newpage

\section{Introduction}

Superstring theory in ten dimensions is a critical
$N=1$ superconformal theory. It can be quantized using a nilpotent
BRST operator 
\begin{equation} 
Q = \frac{1}{2\pi i}\oint dz~
 [c(T_m+\frac{1}{2} T_g) + \gamma(G_m +
\frac{1}{2} G_g)]
\label{qone}
\end{equation} 
where $[T_m,G_m]$ are the $c=15$ N=1 superconformal generators and
$[T_g,
G_g]$ are the $c=-15$ N=1 superconformal generators constructed from 
a pair of fermionic ghosts $[b,c]$ and a pair of bosonic ghosts
$[\beta,\gamma]$. Physical states are described by vertex operators in the
cohomology of $Q$ and, in order to construct vertex operators for the
spacetime fermions, it is convenient to 
fermionize the bosonic ghosts as 
\begin{eqnarray} \beta &=& \partial \xi e^{-\phi} \\ 
\gamma &=& \eta e^{\phi}, 
\label{ferm}
\end{eqnarray} 
where $\eta$ and $\xi$ are free fermions and $\phi$ is a 
chiral boson.\cite{FriedanMartinecShenker}

Because the above
fermionization involves $\partial\xi$ rather than $\xi$, it is not
possible to write the zero mode of $\xi$ in terms of the
[$\beta,\gamma$] ghosts. The Hilbert space without
the $\xi$ zero mode is called the ``small'' Hilbert space, while the
Hilbert space including the $\xi$ zero mode is called the ``large'' Hilbert
space. 
The small Hilbert space can be defined as operators annihilated by
$\oint dz ~
\eta$, and one can show that any such operator can be constructed out 
of the original $[\beta,\gamma]$ ghosts.
Since physical vertex operators should be in the small Hilbert space,
they must be annihilated by both $Q$ and $\oint dz~\eta$. For
consistency, this requires that
$Q$ should not only be nilpotent, 
but should also anti-commute with $\oint dz~\eta$.

In this letter, we will construct a similarity transformation $R$
such that $Q= e^{-R} (\frac{1}{2\pi i}\oint 
dz~ \gamma^2 b) e^R$. (Note that the first
term of $R$ was constructed in \cite{Berkovits}.)
Since $\gamma^2 b$ is nilpotent,
this trivially proves that $Q$ is nilpotent. Furthermore, since
$\oint dz~
\gamma^2 b$ has trivial cohomology, it proves that $Q$ has trivial
cohomology in the large Hilbert space. 

However, $R$ does not
commute with $\oint dz~\eta$, so $Q$ has non-trivial cohomology
in the small Hilbert space as expected. Also, $e^{-R} (\frac{1}{2\pi i}
\oint dz~\gamma^2 b) e^R$
only anti-commutes with $\oint dz~\eta$ in the critical dimension,
so one cannot use the nilpotent
$e^{-R} (\frac{1}{2\pi i}\oint dz~\gamma^2 b) e^R$ to quantize the superstring
when $D\neq 10$.

\section{Similarity transformation}

After fermionizing the $[\beta,\gamma]$ ghosts as in (\ref{ferm})
and bosonizing
$\xi= e^{\chi}$ and $\eta=e^{-\chi}$, the BRST charge 
of (\ref{qone}) can be written as 
$Q = \frac{1}{2\pi i}\oint dz ~
j_{BRST}$ where 
\begin{eqnarray} 
j_{BRST} &=& c[T_m- b\partial c- \partial^2\phi
-\frac{1}{2}(\partial\phi)^2 +\frac{1}{2}
\partial^2 \chi +\frac{1}{2}(\partial\chi)^2)]\nonumber \\ 
&+&  e^{\phi-\chi}G_m 
- be^{2(\phi-\chi)} + \partial^2 c +\partial(c \partial \chi).
\label{Q} 
\end{eqnarray} 
We will now show that 
\begin{equation} 
j_{BRST} = e^{-R} j_0  e^R
\label{simil} 
\end{equation} 
where 
\begin{equation} 
j_0 = -b e^{2(\phi-\chi)},
\label{jzero} 
\end{equation} 
\begin{equation} 
R = \frac{1}{2\pi i}\oint dz~[c G_m e^{-\phi}e^{\chi} - \frac{1}{4}
\partial (e^{-2\phi})e^{2\chi} c \partial c].
\label{R} 
\end{equation} 
Note that 
$j_{BRST}$ was used in \cite{twist}
as the fermionic generator $G^+$ of a twisted 
N=2 superconformal algebra. Using (\ref{simil}), $j_{BRST}$ is
trivially nilpotent since $j_0$ has no poles with itself. 

To prove (\ref{simil}), we use the
expansion
\begin{eqnarray} 
e^{-R} j_0 e^R &=& \sum_{n=0}^\infty \frac{1}{n!}  j_n,\nonumber \\ 
j_n &=& [j_{n-1},R], 
\label{expa} 
\end{eqnarray} 
where, for 
$R=\frac{1}{2\pi i}\oint dz~ r(z) $, 
the commutator is computed following the rule
\begin{equation} 
[j_{n-1}(y),R]=\frac{1}{2\pi i}\oint dz~ j_{n-1}(y)
r(z).
\label{rule} 
\end{equation} 

If $D$ is the spacetime dimension of the superstring (i.e. $G_m(y)
G_m(z) \to D (y-z)^{-3} + ...$), the 
$n=1$ term of (\ref{expa}) is given by 
\begin{eqnarray} 
j_1 &=& e^{\phi}e^{-\chi}G_m + bc\partial c  -
\frac{3}{2} \partial^2 c +
\partial c (5\partial\phi -4\partial\chi)\nonumber \\ 
&+& c[\frac{3}{2}{\partial}^2\phi - 3({\partial\phi})^2 - 2({\partial\chi})^2 -
 \partial^2 \chi + 5\partial\phi\partial\chi], 
\label{j1} 
\end{eqnarray} 
the $n=2$ term is given by 
\begin{eqnarray} 
j_2 &=& 2cT_m + 
\frac{D}{2} [\partial^2 c +
2 \partial c(\partial \chi-\partial\phi)\nonumber \\
&+& c( {\partial}^2\chi -\partial^2\phi + (\partial\phi)^2 +
(\partial\chi)^2 - 2\partial\phi\partial\chi)] \nonumber \\ 
&-& e^{-\phi}e^{\chi} G_m c\partial c
+ \frac{5}{4} e^{-2\phi}e^{2\chi} c\partial c {\partial}^2c, 
\label{j2} 
\end{eqnarray} 
the $n=3$ term is given by 
\begin{equation} 
j_3 = 3e^{-\phi}e^{\chi} G_m c\partial c -
\frac{3D}{4}e^{-2\phi}e^{2\chi} c\partial c {\partial}^2c, 
\label{j3} 
\end{equation}
and the $n=4$ term is given by 
\begin{equation} 
j_4 = \frac{3D}{2} e^{-2\phi}e^{2\chi} c\partial c {\partial}^2c.
\label{j4} 
\end{equation} 
The terms for $n>4$ in the expansion 
vanish identically since the OPE between $j_4$
and $R$ has no single poles.

It is straightforward to check that $j_{BRST}$ of (\ref{Q}) is equal to
\begin{equation} 
j_0 + j_1 +\frac{1}{2!}j_2 + \frac{1}{3!}j_3 + \frac{1}{4!}j_4
\label{jseries} 
\end{equation} 
when $D=10$, so we have proven (\ref{simil}). Note that when $D\neq 10$,
the integral of (\ref{jseries}) contains the term 
$(10-D) \oint dz (\frac{1}{2} c \partial\phi\partial\chi+
\frac{1}{16} e^{2(\chi-\phi)}c\partial c
\partial^2 c)$. 
Since this term does not anti-commute
with $\oint dz~
\eta =\oint dz ~
e^{-\chi}$, the integral of (\ref{jseries})
can only be used as a 
BRST charge when $D=10$ for the reasons stated in the introduction.

\section {Acknowledgements}

JNA would like to acknowledge financial support from FAPESP grant 
number 98/12362-2, 
NB would like to acknowledge partial financial support from CNPq
grant number 300256/94-9, and OCH would like to acknowledge
financial support from FAPESP grant number 98/02380-3.

\end{document}